\RequirePackage{fix-cm}

\documentclass[smallcondensed,12pt]{svjour3}     
\smartqed  
\usepackage{graphicx}
\usepackage{natbib}

\begin{document}

\title{Coorbital capture at arbitrary inclination}



\author{Fathi Namouni    \and
        Helena Morais }


\institute{F. Namouni \at
              Universit\'e C\^ote d'Azur, CNRS, Observatoire de la C\^ote d'Azur, CS 34229, 06304 Nice, France\\
                           \email{namouni@obs-nice.fr}           
           \and
           M.H.M. Morais\at
             Instituto de Geoci\^encias e Ci\^encias Exatas, Universidade Estadual Paulista (UNESP), Av. 24-A, 1515 13506-900 Rio Claro, SP, Brazil\\ \email{helena.morais@rc.unesp.br}
}

\date{Received: date / Accepted: date}

\maketitle

\begin{abstract}
The process of capture in the coorbital region of a solar system planet is studied. Absolute capture likelihood in the 1:1 resonance is determined by randomly constructed statistical ensembles numbering $7.24\times 10^5$  of massless asteroids that are set to migrate radially from the outer to the inner  boundaries of the coorbital region of a Jupiter-mass planet. Orbital states include coorbital capture,  ejection, collisions with the Sun and the planet and free-crossing of the coorbital region. The relative efficiency of retrograde capture with respect to prograde capture is confirmed as an intrinsic property of the coorbital resonance. Half the asteroids cross the coorbital region regardless of eccentricity and for any inclination less than $120^\circ$.  We also find that the recently discovered retrograde coorbital of Jupiter, asteroid 2015 BZ509, lies almost exactly at the capture efficiency peak associated with its orbital parameters. 

\end{abstract}

\section{Introduction}
\label{intro}
\label{sec:1}
The recent discovery of asteroid 2015 BZ509 with a retrograde orbit inside Jupiter's coorbital region has renewed interest in the dynamics of the coorbital 1:1 resonance as its peculiar configuration does not seem at first sight in accord with its long lifetime of at least one million years \citep{Wiegert17,MoraisNamouni17}. Asteroid 2015 BZ509 in the 1:1 resonance with Jupiter is not the first of its kind to share the orbit of a solar system planet. In our study of  the dynamics of Centaurs and Damocloids \citep{MoraisNamouni13b}, several  bodies were found captured in retrograde mean motion resonances with Jupiter and Saturn. One in particular, 2006 BZ8, will be captured in the coorbital region of Saturn in the near future. These observations motivated us to study the process of mean motion resonance capture at arbitrary inclinations in the solar system. Using large scale statistical  simulations  \citep{NamouniMorais15,NamouniMorais17}, we studied the three-body system, Sun-Jupiter-asteroid, when the small object  is placed outside the outer 1:5 resonance ($\sim 15$\,AU) and subjected to a slow radial drift towards the inner solar system thereby encountering the planet's complex web of mean motion resonances. The asteroid was allowed to travel the system until it reached  $\sim 1.2$\,AU. The thirty main mean motion resonances were examined and capture probabilities measured among which that of the coorbital 1:1 resonance. One of the main result of these analyses was the unexpected large efficiency of retrograde resonance capture  with respect to that of prograde resonance capture. In particular, the coorbital 1:1 resonance which is the strongest mean motion resonance for both prograde and retrograde motion was found to have near unit capture probability for the most retrograde orbits. Another interesting result was the presence of a dynamic corridor past the planet (and hence past the coorbital resonance) that allowed astroids to travel way inside the planet's orbit and be captured in high order inner resonances in great numbers \citep{NamouniMorais17}. Since asteroids crossed so many resonances in their journey towards the Sun in the previous works' setting, the capture probability of a given resonance (with the exception of the outer 1:5 that was first to be encountered)  is conditional and not absolute thus preventing us from gauging the intrinsic workings of coorbital capture. It is the object of this work to shed some light on the intrinsic capture dynamics of the coorbital 1:1 resonance for arbitrary inclinations using the same tools as the two previous works in order to facilitate their comparison. 

In order to understand the workings of the coorbital 1:1 resonance, we will make use of a number of results from previous works on the subject. It is useful to remind the reader of some of the dynamical features peculiar to the 1:1 resonance. Orbits in the coorbital region come in various types the simplest of which for prograde motion are tadpoles (T), horseshoes (H)  and retrograde satellite (RS) orbits that correspond to  librations of the critical argument $\phi=\lambda-\lambda_p$ around $\pm60^\circ$, $180^\circ$ and $0^\circ$ respectively where $\lambda$ and $\lambda_p$ are the longitudes of the asteroid and the planet  \citep{ssdbook}. Such orbital types may occur when the eccentricity and inclination are moderate. For larger values of both eccentricity and inclination, the basic orbits may merge forming  the so-called compound orbits \citep{Namouni99,Namounietal99} (such as H-RS, T-RS and RS-T-RS) and the 1:1 resonance libration becomes a complex process intertwined with stable periodic orbital transitions between the different types of motion and the Kozai-Lidov resonances (hereafter KL-resonances) \citep{Kozai62,Lidov62,Namouni99}. In partcular, the KL-resonance may maintain the astroid in the Kozai resonance without libration of the argument $\phi$. Such motion was termed a passing orbit (or P-orbits) of the coorbital region. When motion is retrograde, the simplest orbital types involve librations of the critical angles $\phi$ and $\phi^\star=\phi-2\omega$ where $\omega$ is the argument of perihelion. These  retrograde motion types have been denoted R1 when $\phi^\star$ librates around around $0^\circ$, R2 and R3 when   $\phi^\star$ librates $180^\circ$   at  large and small eccentricities respectively,  and R4 when $\phi$ librates around $180^\circ$  \citep{MoraisNamouni13a,MoraisNamouni16}. The latter mode occurs only in three dimensions. 

The article is structured as follows:  in Sec. \ref{sec:2}, the numerical setup of the coorbital capture simulations is discussed. In Sec, \ref{sec:3}, the main four simulations outcomes (long-lived capture, ejection, solar collisions and crossing orbits) are presented and their relative likelihoods are discussed. In particular, we also confront the absolute likelihood of coorbital capture derived in the work to the conditional likelihood derived in our previous works. Sec. \ref{sec:4} contains our concluding remarks.

\section{Simulation setup}
\label{sec:2}
In order to assess the absolute capture likelihhod of the coorbital resonance, we consider  the three-body problem of  \citep{NamouniMorais15,NamouniMorais17} where a Jupiter mass planet on a circular orbit of semi-major axis $a_p$ revolves around a solar-mass.  A massless asteroid with  semi-major axis $a_0=1.1 a_p$ outside the  planet's coorbital region of half-width the Hill radius $R_H=0.07 a_p$. An imposed slow inward semi-major axis drift will make the particle's orbit cross the coorbital region. Drift is modeled through a velocity-dependent drag force of the form $-k {\bf v}$ leading to  a semi-major axis function $a(t)=a_0\exp(-2k t)$.  This drag force does not model a specific physical process that the asteoids may be subjected to, it is used to study resonance crossing with the simplest deterministic and unidirectional radial drift that does influence only the semi-major axis evolution and leaves the other orbital elements such as eccentricity and inclination unchanged. We choose a long characteristic drift times are used $\tau=(2k)^{-1}=10^7 T_p$ where $T_p$ is the planet's  orbital period.  In terms of physical units: $a_p=5.2 $\, AU  and the drift time is $1.2 \times 10^8$ years.  The asteroid's inclination $I_0$ is fixed as an initial condition and is varied from $1^\circ$ to $179^\circ$ with a $1^\circ$-inclination step. Two additional values $I_0=0.001^\circ$ and $I_0=179.999^\circ$ are used to complete the inclination range. The boundary values $0^\circ$ and $180^\circ$ are not used because the dynamics of exactly two-dimensional capture differ significantly from three-dimensional capture \citep{MoraisNamouni16} and the capture probabilities that interest us are those that may explain the presence of real asteroids in Jupiter's coorbital region such as 2015 BZ509. 

For each inclination and each eccentricity standard deviation  $\sigma_e=0.01$, $0.1$, $0.3$ and $0.7$, we generate  a statistical ensemble of 1000 particles. The probability distributions of the eccentricity $e$ and longitude of perihelion $\varpi$ are obtained by applying a two-dimensional normal distribution of the eccentricity vector ($e\cos \varpi$,$e\sin\varpi$). This implies a Rayleigh distribution for the eccentricity amplitude $e$. Mean longitudes $\lambda$ and longitudes of the ascending nodes $\Omega$ are taken from uniform distributions.  The study totals $724\,000$ numerical simulations of resonant capture. 

In order to be able to compare the outcomes of all simulations and assess absolute probability likelihood, we require that the simulations stop after a time equal to that which takes a particle unperturbed by the planet to cross the coorbital region and reach the semimajor axis $a_f=0.9a_p$, the symmetrical value of $a_0$ with respect to the planet's orbit. By applying the drift law, we get the simulation time  $T_s=2\times 10^6 T_p$. As we explain in the next section, the fixing of physical boundaries ($0.9a_0$ and $1.1a_0$) gives a specific meaning to the state of capture in resonance as that of long-lived capture.

\section{Simulation outcomes}
\label{sec:3}
Five  orbital outcomes are observed in the simulation: captured asteroids, ejected astroids, astroids that collide with the star, those that collide with the planet and capture-resistant asteroids that reach the inner physical boundary at $0.9 a_p$ that we shall call crossing asteroids.  

In our particular setting where the physical boundaries translate into a fixed integration timespan, capture in the coorbital region means that the asteroid survives  the full simulation duration. It is important to emphasize that asteroids may be captured for shorter times after which they may be ejected or pass across the planet's coorbital region towards smaller semi-majors axes. Such temporary capture states are not added to the capture tally  as we are aim to understand long-lived capture. We also note that unlike the case of high order resonance capture where asteroids were ejected because of close encounters with the planet while they were in resonance \citep{NamouniMorais17}, we observe that most asteroids ejected near the coorbital resonance have random walked the outer boundary of the coorbital region (from where they were placed) and eventually experienced close encounters with the planets without having been captured in resonance. Ejection is thus mainly  caused by the chaotic layer  outside the coorbital region generated by the overlapping of mean motions resonances of the type $p$:$p+1$ with $p\gg 1$ \citep{Wisdom80}. 

\begin{figure*}
\includegraphics[width=0.63\textwidth]{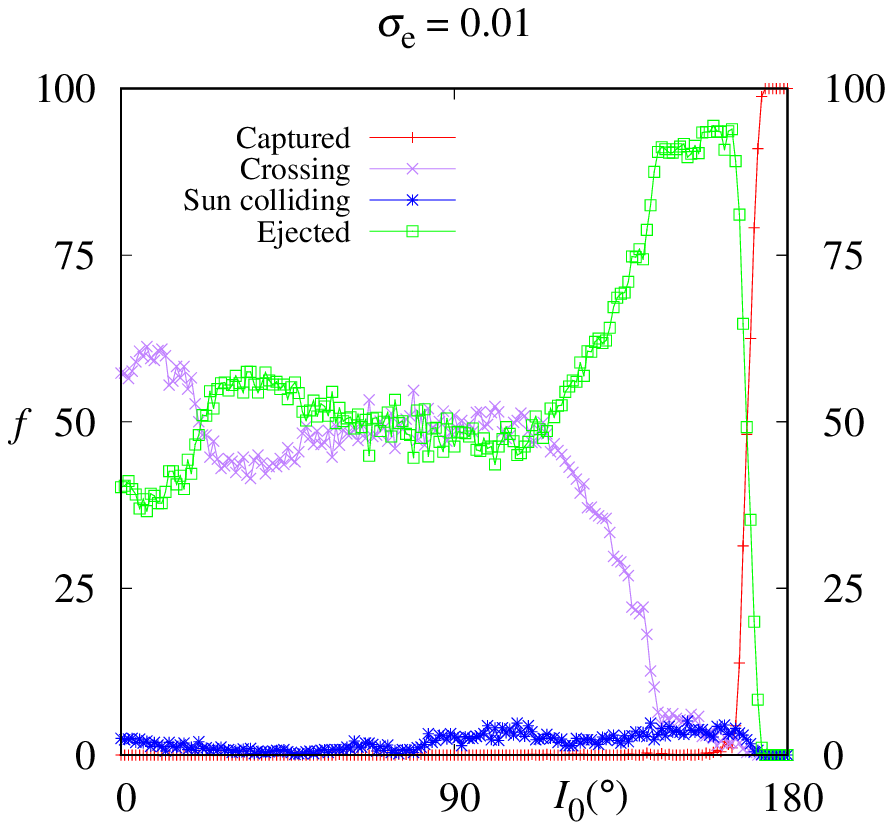}\hspace{-21mm}\includegraphics[width=0.63\textwidth]{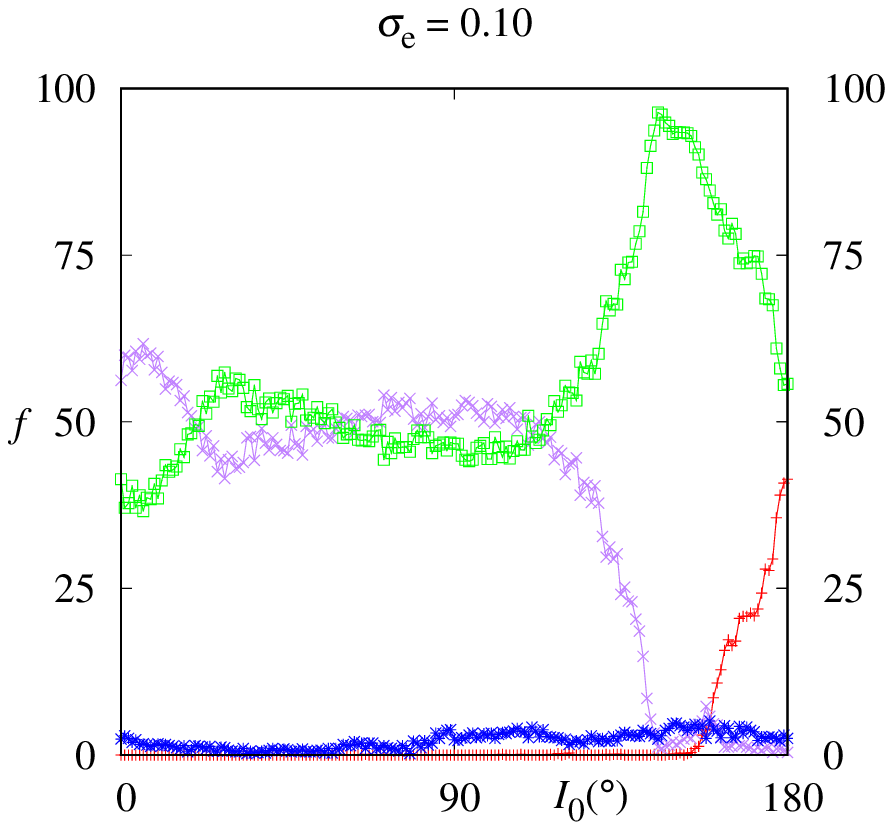}\\
\includegraphics[width=0.63\textwidth]{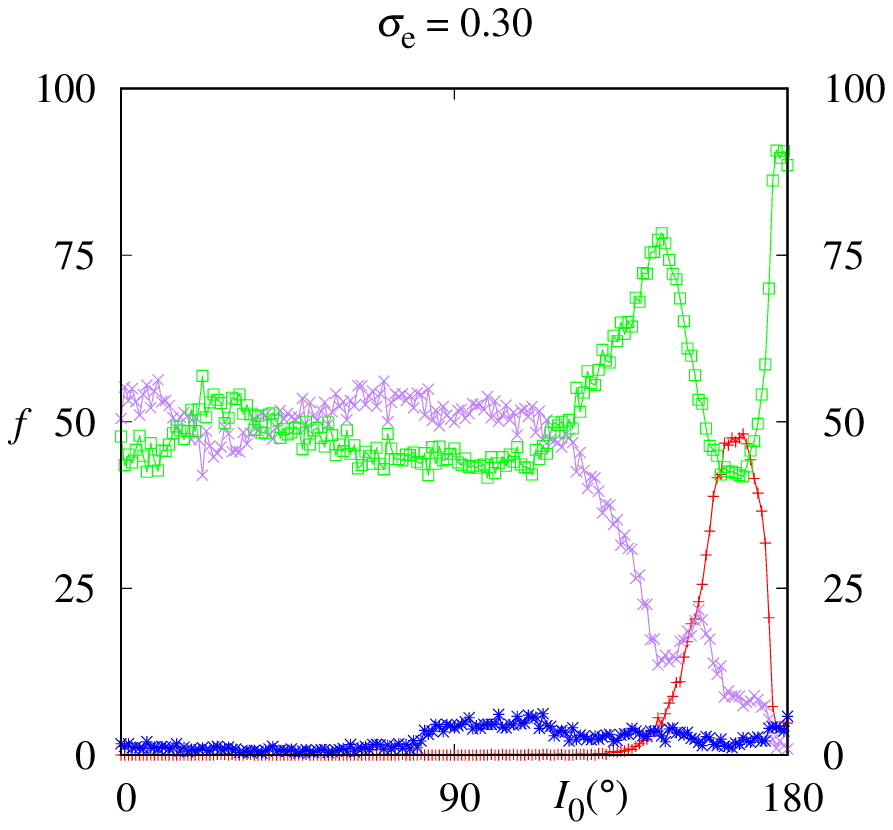}\hspace{-21mm}\includegraphics[width=0.63\textwidth]{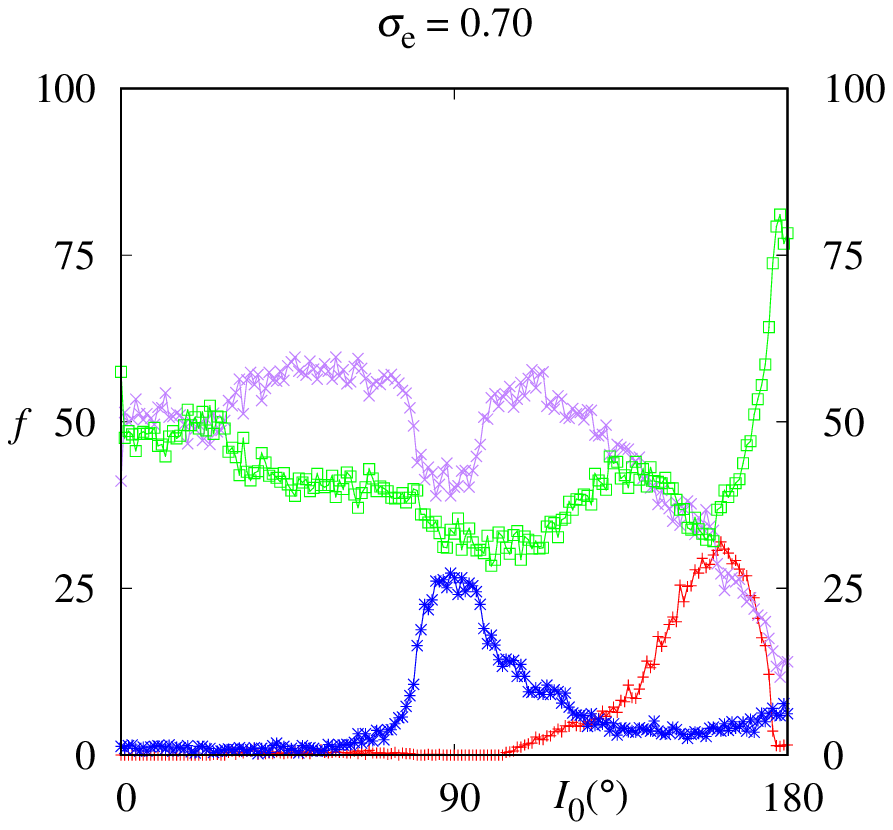}\\
\caption{Fraction of asteroids $f(\%)$ in a given state (long-lived capture, ejection, Sun-collision, crossing) as a function of initial inclination $I_0(^\circ)$ for the four eccentricity standard deviations $\sigma_e$.}
\label{fig:1}    
\end{figure*}

In Fig. \ref{fig:1}, we show the absolute probability for the simulation outcomes in terms of  the ratio $f$  in \% of the number of asteroids in  a given state to the total number of particles (1000) in the sample as a function of initial inclination $I_0$ for each eccentricity standard deviation $\sigma_e$.  It is important to bear in mind that since our samples for a given $I_0$ and $\sigma_e$ contain 1000 asteroids, our capture probability is known at best to 0.1\%. We note that despite the large number of ejected particles, collisions with the planet are rare and are therefore not indicated in the Figure.

We find that for small eccentricity orbits ($\sigma_e=0.01$), capture is 100\% certain if $I_0\geq 174^\circ$ and is virtually rare for $I_0<150^\circ$. Ejection peaks at $f=90\%$ for $145<I_0<165^\circ$, flattens out around $f=55\%$ until $I_0\sim 55^\circ$ and decreases slightly below that inclination towards 40\%. The ejection probability is almost exactly complementary to the crossing orbits' probability for all inclination $I_0<165^\circ$ where the capture probability  all but vanishes. Collisions with the Sun occur with a small probability $f\sim 4\%$ for most retrograde orbits and prograde ones with  $I_0\leq 40^\circ$.  As the eccentricity is increased to $\sigma_e=0.1$, the previous trends remain albeit with a capture probability reduced by more than half. For $\sigma_e=0.3$, capture at nearly coplanar orbits is not longer certain near $I=180^\circ$ ($f=5\%$) and the efficiency peak is displaced to $I_0=166^\circ$ and $f\sim 50\%$. Interestingly, asteroid 2015 BZ509 lies almost exactly at the maximum efficiency peak (the orbit's eccentricity and inclination are $e=0.38$ and $I=163^\circ$).  Ejection peaks at ($f=90\%$,$I_0=178^\circ$) and ($f=78\%$,$I_0=145^\circ$). Below the inclination $I_0=155^\circ$, crossing orbits and ejected orbits probabilities are  complementary and flatten out to $f=50\%$. Collisions with the Sun are marginally increased for retrograde orbits and decreased for prograde orbits and there appears to be a certain preference polar orbits. For the largest eccentricity ensembles, $\sigma_e=0.7$, the capture efficiency peak occurs at slightly lower inclination at ($f=30\%$,$I_0=162^\circ$) and  the probability width is significantly increased towards nearly polar orbits. Ejection dominates capture and crossing orbits from nearly coplanar retrograde inclinations down to the capture efficiency peak where crossing orbits become the likeliest outcome. Near $I_0\sim180^\circ$, crossing orbits at $f=14\%$ and sun-colliding orbits at $f=7\%$ dominate captured orbits  whose $f=1\%$. The Sun-collision probability peaks for initially polar orbits at $f=27\%$ and has a width of $20^\circ$ at half-maximum displaced towards retrograde orbits. The probability increase of sun-colliding orbits is matched by a net decrease of the crossing orbits' probability. This may be explained by the geometry of high eccentricity polar orbits that once captured in resonance, increase their eccentricity without the interference of close planet encounters and eventually hit the star as $e$ approaches unity. One of the interesting outcomes of these simulations is the significant probability of crossing asteroids regardless of any initial eccentricity and any initial inclination less than $120^\circ$. With $f\sim 50\%$ this explains how the dynamical corridor that was observed in our earlier work  is possible for subpolar orbits past the planet's coorbital region \citep{NamouniMorais17}.

\begin{figure*}
\includegraphics[width=0.63\textwidth]{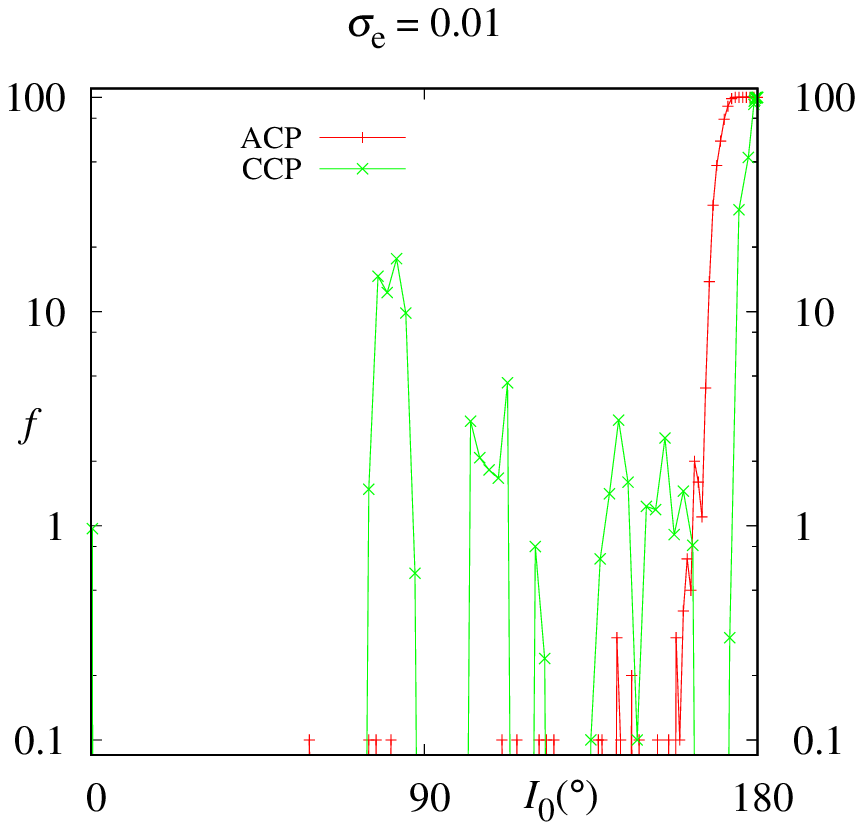}\hspace{-21mm}\includegraphics[width=0.63\textwidth]{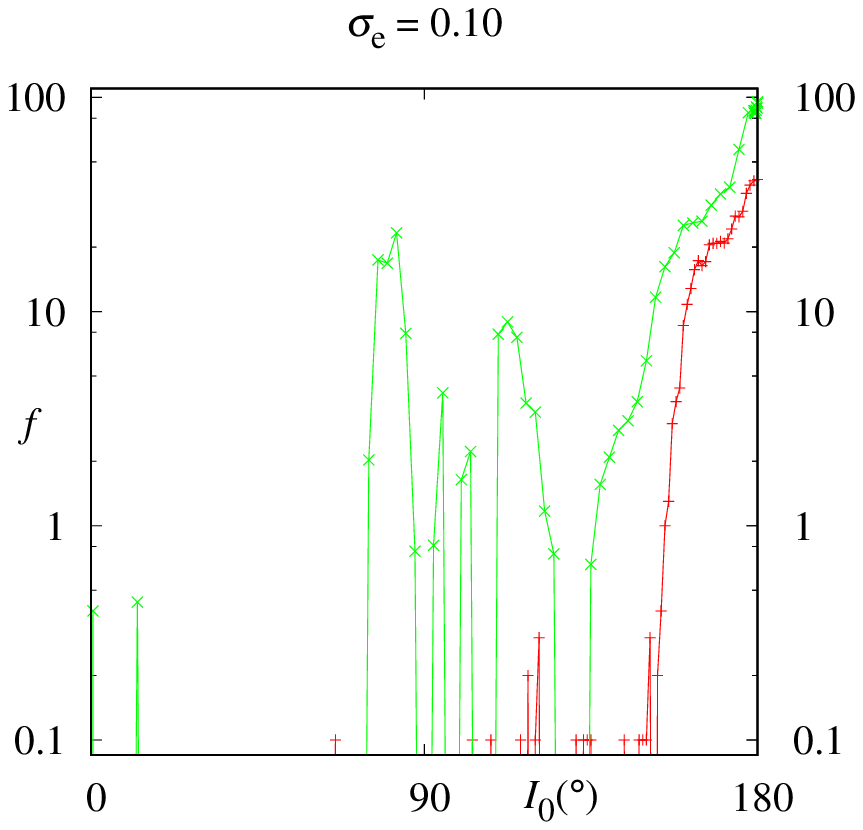}\\
\includegraphics[width=0.63\textwidth]{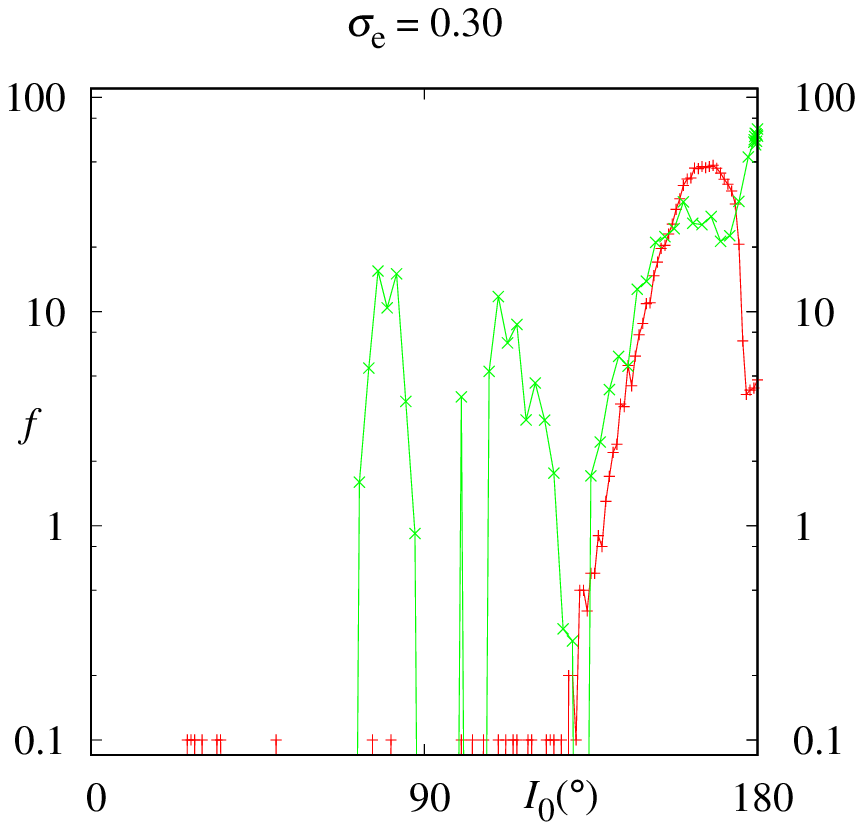}\hspace{-21mm}\includegraphics[width=0.63\textwidth]{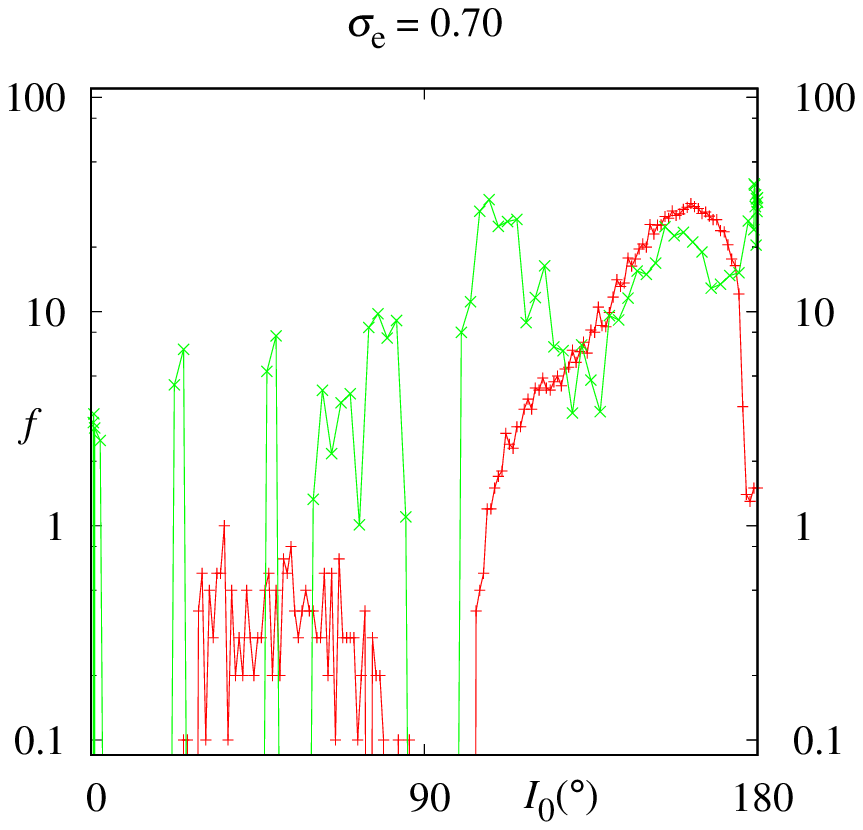}\\
\caption{Comparing the absolute capture probability  to the conditional capture probability  of  \citep{NamouniMorais17} as a function of initial inclination $I_0(^\circ)$ for the four eccentricity standard deviations $\sigma_e$.}
\label{fig:2}      
\end{figure*}

In our earlier works on resonance capture at arbitrary inclination, we set up the asteroids outside the location of the 1:5 resonance with the Jupiter-mass planet at $\sim 15$\, AU and let them radially migrate  with the same drift force of the present simulations  for a timespan that allows an unperturbed asteroid to reach $\sim 0.2$\,AU. Some of these asteroids were captured by the planet's 1:1 coorbital resonance. The derived capture probability is not absolute as the captured bodies crossed a large resonance web that modified their initial inclinations and eccentricities \citep{NamouniMorais15,NamouniMorais17}. In Fig. \ref{fig:2} we compare the absolute capture probability (hereafter ACP) derived in this work and shown in Fig. \ref{fig:1} with the conditional one (hereafter CCP) measured in  \citep{NamouniMorais17}. We present the conditional probability corresponding to the same drift time $10^7T_p$ and choose a logarithmic probability scale so that we may assess the similarities and differences for all capture likelihoods including those with very small amplitudes that were not clearly visible in Fig. \ref{fig:1}. In particular, for $\sigma_e=0.70$, we now may see clearly  the capture likelihood  $0.1\leq f<1\%$ of the ACP inclination domain $30^\circ\leq I_0\leq 80^\circ$.  We also remind the reader that the smallest probability value  in both works is 0.1\% on account of the simulations' statistical resolution as explained above so that below that value the probability is zero. Two salient features are noted in Fig. \ref{fig:2}: first is the fact that the probability distributions are similar for the most retrograde orbits for small ($\sigma_e=0.01$) to moderate eccentricity ($\sigma_e=0.1$) although for the former ensembles, ACP is wider and larger than  CCP whereas for  the latter ensembles, the trend is inverted. For medium to large eccentricities ($\sigma_e=0.1$ and 0.7), ACP and CCP are almost identical for the most retrograde orbits except near $I_0= 180^\circ$. The second main feature is the presence of additional probability peaks located for the most part near  inclinations $I_0$ where ACL is smallest reflecting the fact that the CCP orbits have evolved and their eccentricities were significantly modified
before their arrival at the boundary of the coorbital region. Most prominent in the probability distributions are the sub-polar and super-polar peaks on both sides of $I_0=90^\circ$ where capture likelihood reaches 10\% to 30\%. There are however additional isolated CCP peaks notably for nearly coplanar prograde orbits where ACL is zero. This indicates that the corresponding orbits' inclinations were significantly modified  to allow capture. 
 
\section{Conclusions}
\label{sec:4}
In this work, we studied the absolute capture probability of the coorbital 1:1 resonance using statistical simulations that allow us to meaningfully measure the number of captured asteroids as a function of the initial inclinations for four initial eccentricity standard deviations. Our main findings are: (1)  the efficiency of coorbital capture at large retrograde inclination is confirmed to be an intrinsic feature of coorbital resonance, (2)  the dynamical corridors observed in \citep{NamouniMorais17} are facilitated by the fact that half the asteroids that try and cross the coorbital region are not capture and allowed safe passage, (3) upon arriving near the coorbital region, asteroids may be ejected not by the coorbital resonance but by the chaotic layer that surrounds it, (4) Jupiter's retrograde coorbital 2015 BZ509 sits near the peak of coorbital capture efficiency. Our next step in the study of resonance capture  at arbitrary inclination is to uncover the origin of retrograde asteroids and  Centaurs that  wander safely among the solar system's outer planets and get captured in the coorbital resonance.

\begin{acknowledgements}
F. N. thanks the 2016 Col\'oquio Brasileiro de Din\^amica Orbital Organizing Committee for their kind invitation to the conference where part of this work was presented. The authors acknowledge support from grant 2015/17962-5 of S\~ao Paulo Research Foundation (FAPESP). The numerical simulations in this work were performed at the Centre for Intensive Computing  `M\'esocentre {\sc sigamm}'  hosted by the Observatoire de la C\^ote dÕAzur.

\end{acknowledgements}

\bibliographystyle{spbasic}      
\bibliography{ms}   

\end{document}